\documentstyle[12pt,aaspp4]{article}

\lefthead{Nilsson et al.}
\righthead{Optical jet in 3C 371}

\begin{document}

\title{
Discovery of an Optical Jet in the BL Lac Object 3C 371
\altaffilmark{1,2}}

\author{K. Nilsson}
\affil{Tuorla Observatory, V\"ais\"al\"antie 20, FIN-21500 Piikki\"o, Finland}

\author{J. Heidt}
\affil{Landessternwarte Heidelberg, K\"onigstuhl, D-69117 Heidelberg,
Germany}

\author{T. Pursimo, A. Sillanp\"a\"a and L. O. Takalo}
\affil{Tuorla Observatory, V\"ais\"al\"antie 20, FIN-21500 Piikki\"o, Finland}

\author{K. J\"ager}
\affil{Universit\"atssternwarte G\"ottingen, Geismarlandstr. 11,
D-37083 G\"ottingen, Germany}

\altaffiltext{1}{Based on observations carried out with the Nordic
Optical Telescope, La Palma, Canary Islands}
\altaffiltext{2}{Based on observations collected at the German-Spanish
Astronomical Centre, Calar Alto, operated by the Max-Planck-Institut
f\"ur Astronomie, Heidelberg, jointly with the Spanish National 
Commission for Astronomy}

\begin{abstract}
We have detected an optical jet in the BL Lac object 3C 371 that 
coincides with the radio jet in this object in the central few kpc.
The most notable feature is a bright optical
knot 3\arcsec\ (4 kpc) from the nucleus that occurs at the location where
the jet apparently changes its direction by $\sim$ 30\arcdeg.
The radio, near-infrared  and optical observations of this knot are 
consistent with a single power-law spectrum with a 
radio-optical spectral index
$\alpha_{ro}$ = -0.81. One possible scenario for the observed turn
is that the jet is interacting with the material in the bridge
connecting 3C 371 to nearby galaxies and the pressure gradient is
deflecting the jet significantly. 
\end{abstract}


\section{Introduction}

The number of known optical counterparts of radio jets in radio
galaxies and quasars has been steadily increasing during the last 
few years. At least seven such cases are now known (M 87, 3C 273,
PKS 0521-36, 3C 66B, 3C 264, 3C 78 and 3C 120) and even more
objects are suspected to harbour an optical jet
(e.g. de Koff et al. \cite{kof:bau}). In one case,
3C 66B (Fraix-Burnet \cite{fra:fra}), optical emission has been
tentatively detected also in the counterjet side.
The radio emission of the jets is thought to be syncrotron radiation
from relativistic electrons spiralling in the magnetic field of the
jet and transporting energy to the outer lobes. At some high frequency
corresponding to the highest energy electrons the spectrum cuts off
and the intensity falls quickly. If this cutoff frequency
is high enough, optical emission can be detected from the jet.

As the number of confirmed optical jets has increased, some patterns
in their properties have emerged. In optical the jets always seem to have
a ``knotty'' appearance and their structure in radio and optical is
very similar. However, small differences in the optical and radio 
structures can be seen. For instance, in a detailed study of M 87,  
Sparks, Biretta \& Macchetto (\cite{spa:bir}) found that the optical/UV 
emission seems to be more 
concentrated in the central parts of the jet and the knots show
higher contrast in the optical than in the radio. Also, the radio-optical
spectral index seems to be flatter in the knots than in the interknot 
regions. The optical jets are usually small in comparison with the
radio jets that can continue tens of kiloparsecs away from the
nucleus. Some optical jets display twists and/or bifurcations like
the jet in 3C 264 (Crane et al. \cite{cra:pel}).

The apparent onesideness of the optical jets indicates that 
relativistic beaming is playing an important role in modifying the
intrinsic intensity distribution. If the sources with optical jets
have their jets directed close to the line of sight, then these 
sources should be smaller than corresponding radio sources on average. 
Sparks et al. (\cite{spa:gol}) compared the linear sizes of a
sample of 3CR radio galaxies with optical jets to those without a jet 
and found the radio structures of the galaxies with an optical jet to 
be smaller by a factor of $\sim$ 3. This is in accordance with the 
relativistic beaming picture,
but intrinsic differences between the jet and counterjet cannot be
ruled out yet. The optical counterjet in 3C 66B (Fraix-Burnet
\cite{fra:fra}), if confirmed, would argue for intrinsic differences
between the jet and counterjet as the optical emission in the
counterjet side of 3C 66B is much stronger than what is expected from
relativistic beaming. 

  
3C 371 is a nearby (z=0.051) BL Lac object that belongs to the 1 Jy
BL Lac sample of Stickel et al. (\cite{sti:pad}). The BL Lac object is
hosted by a luminous (M$_V$ = -22.4) elliptical galaxy that is
surrounded by a small cluster of galaxies (Stickel, Fried \& K\"uhr
\cite{sti:fri} and references therein). It has been mapped with the
VLA at 6 cm by Wrobel \& Lind (\cite{wro:lin}) 
who found a dominant central core and two faint lobes east and west 
of the core. The eastern lobe is very diffuse with no apparent
connection with the nucleus, but the western lobe is connected to the 
nucleus by a wiggling jet that terminates into a weak hotspot inside
the lobe (see Fig. 4 [Plate 2]). Wrobel \& Lind (\cite{wro:lin})
interpret this structure as an edge 
brightened double radio source with the angle between the source axis 
and line of sight large enough so that the lobes do not overlap.
At the distance of $\sim$ 3 arcsec from the nucleus the jet
apparently turns $\sim$ 30\arcdeg. At this location a sudden
brightening of the jet is seen. This bright knot has been resolved by
Akujor et al. (\cite{aku:lud}) at 18 cm with the MERLIN array.
Their 0\farcs25 resolution map shows the bright knot to be extended
in the direction roughly perpendicular to the line connecting the knot and
the nucleus (see Fig. 2).  

\section{Observations and data analysis}

The R and B-band observations
were carried out at the 2.56m Nordic Optical Telescope (Table 1).
The pixel scales of the TEK and Loral chips were 0\farcs176 pix
$^{-1}$ and 0\farcs11 pix $^{-1}$, respectively. The R-band
observations of 20 Oct 1995 consist of 6 $\times$ 600s exposures
that have saturated the nucleus of the galaxy. In 13 Jul 1996
the individual exposures were kept between 120 and 180 seconds to
avoid saturation. The B-band observations consist of 6 $\times$ 600s
exposures with no saturation of the nucleus. The images were de-biased
and flat-fielded with IRAF using twilight flats and the individual 
exposures were registered and averaged. Photometric calibration
of the field was achieved on the night of 17 Jun 1996 by observing
standard stars from Landolt (\cite{lan:la1},\cite{lan:la2}).
The accuracy of this calibration is $\sim$ 0.04 mag in R and 
$\sim$ 0.06 mag in B.
As a check of our photometry we measured a star in the 3C 371
field that is listed in the UBV calibration sequence of McGimsey 
\& Miller (\cite{gim:mil}). For star 4 in their sequence (B = 16.98
$\pm$ 0.04) we measure B = 16.97 $\pm$ 0.06. This star was also 
used for photometric calibration of the R-band observations on 
20 Oct 1995.

The K$'$-band observations were 
carried out at the Calar Alto 3.5m telescope using the MAGIC NIR
camera. The pixel scale was 0\farcs32 pix $^{-1}$. A total of 60 
frames were obtained, each consisting of 15 frames of 2 seconds each
which were summed and saved. For background subtraction the exposures 
on target were interlaced with exposures of a field a few arcmin away 
from 3C 371. A total of 16 such exposured were obtained, each 15 $\times$ 2
seconds. The science exposures were background subtracted by the 
scaled median of the skyframes and scattered light and bad pixels were
removed. After 
flat-fielding with domeflats the individual frames were registered
and coadded following McLeod \& Rieke (\cite{leo:rie}). Standard stars
from Elias et al. (\cite{eli:fro}) were observed for photomeric 
calibration. Although
there were some thin clouds during the night, the standard stars
gave consistent results and we believe the accuracy to be much
better than 0.1 mag. 

To remove the light from the the host galaxy and core component
in the B and R-band images we first determined the PSF from field
stars and subtracted a scaled
PSF at the location of the nucleus. The scaling was such that a
maximum amount of light was subtracted without producing a ``hole''
in the underlying galaxy. The subtraction was made to facilitate later
analysis by removing the diffraction spikes around the nucleus.
The ISOPHOTE package in IRAF was then
used to fit ellipses to the galaxy light distribution and a model
galaxy image was created and subtracted from the original image. 
This process was iterated a few times to find and mask objects
that were overlapping the 3C 371 host galaxy. The resulting residual
images (Fig. 1 a,c [Plate 1]) have a background very close to zero
showing that the galaxy light distribution has been succesfully modelled.
Outside the nuclear region (r $>$ 4\arcsec) the intensity profile first
follows de Vaucouleurs profile very well, until at a radius of $\sim$
13\arcsec\ (18 kpc) the profile starts to deviate from this law in
the form of excess emission (see Fig. 4).

In the K$'$ band image we do not have a bright star in the field to
model and subtract the diffraction spikes. We thus proceeded directly 
with ellipse fitting by masking overlapping objects as previously and 
rejecting a fraction (8\%) of highest pixels
from the fit. A dark ring around the nucleus and four diffraction
spikes are visible in the residual image (Fig. 1e).
Also, a relatively bright object is seen $\sim$ 2\arcsec\ from
the nucleus at PA $\sim$ 160\arcdeg. After careful analysis of
the individual K$'$-band images we concluded that
this feature is caused by telescope movements during the 15 $\times$
2 second exposures (only the sum of the 15 exposures was saved). 
We estimate that due to this effect about 6\% of the light is
moved from the center of 3C 371 to the feature 2\arcsec\ away.
This also affects the K$'$-band photometry, but we have not made any 
correction for this since we are not certain of the exact amount of
correction needed.

For the comparison of radio and optical images we measured the
positions of 10 stars in the R-band field that can be found in the APM
Northern Sky Catalogue (Irwin, Maddox \& McMahon \cite{irw:mad}) and 
determined the transformation from pixel positions to equatorial coordinates.
Our astrometry indicates that the positions of the optical and radio 
nuclei agree within the positional errors that we estimate to be 
0\farcs8.

\section{Results and discussion}

Fig 1. shows a  19\arcsec $\times$ 19\arcsec\ field around the nucleus
of 3C 371 in all three bands after removing the core and galaxy light from
the images. The left panels show the ``raw'' images and the right
panels show the corresponding fields after 40 deconvolutions with 
the Lucy algorithm (Lucy \cite{luc:luc}). Several objects can
be seen to the west of the nucleus. There is clearly optical and NIR
emission coming from the point where the jet apparently changes its 
direction and where a bright radio knot is also seen. This can be seen
in Fig. 2 where we have overlaid the deconvolved best-resolution
R-band image of 13 Jul 1996 with the 
18 cm radio contours from Akujor et al. (\cite{aku:lud}). The knot
is at a distance of 3\arcsec\ (4 kpc, H$_0$ = 50 km s$^{-1}$
Mpc$^{-1}$) from the nucleus at PA 240\arcdeg\ and
it shows similar structure in optical as in the radio, i.e. it is
extended in the direction roughly perpendicular to the jet direction. 
Some faint emission trailing the knot can also be seen, most notably
in the B-band image. The sharp features around the nucleus in the
deconvolved R and B-band images are probably artifacts caused by
incomplete removal of diffraction spikes. Two faint objects are
visible further out from the knot in the direction where the radio jet
continues in radio maps. The one furthest out from the nucleus is
consistently seen in B and R-band images at the same location. As
there is no one-to-one correspondence between the radio and optical 
emission here, their relation to the radio jet is unclear. The outer
one of these two faint objects could be one of the numerous background
objects embedded in the light from 3C 371 host and the inner one
is probably a noise peak amplified by the deconvolution process. 

Figure 3 shows the broadband spectrum of the 3\arcsec\ knot. The
radio measurements were obtained from the literature and the optical and
NIR data points are from our observations. X-ray data are also
available in the literature (e.g. Comastri, Molendi \& Ghisellini
\cite{com:mol}) but they are not employed here since they always refer
to the total emission from 3C 371. Table 2 lists the results
of our photometry. Here we have converted the K$'$-band flux
to K-band flux using the formula by Wainscoat \& Cowie
(\cite{wai:cow}). The R-band flux is the average of
the observations on 20 Oct 1995 and 13 Jul 1996. The errorbars in
the radio are either from the original publication or if no error 
was given a 1$\sigma$ error
of 10\% was assumed. In our measurements the errorbars include the
errors from photon statistics and calibration, but do not include the
possible error from incomplete background subtraction at the location
of the objects.  The radio-optical
spectral index $\alpha_{ro}$ of the knot is -0.81 (S$_{\nu} \propto
\nu^{\alpha}$) which is within the range of observed 
radio-optical spectral indices in jets, typically -0.9 -- -0.6
(Crane et al. \cite{cra:pel}).
The spectral index in the radio domain is
$\alpha_r$ = -0.70 (or -0.72 if one ignores the deviant point at log
$\nu$ = 9.2). Thus some steepening of the spectrum on the optical
is indicated, but given the errors in Fig. 3 the observations
can well be fitted with a single power-law from the radio to the
optical with no indication of a break or cutoff in the spectrum.  

The jet and overall radio morphology of 3C 371 bear some resemblance
to those in M 87. The jet
in M 87 is relatively faint until an apparent distance of $\sim$
12\farcs5 where a sudden brightening (knot A) is seen both in radio and
optical/UV regions (e.g. Sparks, Biretta \& Macchetto
\cite{spa:bir}). Further out
from the nucleus faint concentrations of emission are seen until the
jet fades from visibility at a distance of $\sim$ 20\arcsec.
The lifetimes of the
electrons emitting optical synchrotron radiation are generally
too short for them to be supplied by the nucleus ( $\sim$ few hundred
years in the case of M 87, see e.g. Biretta, Stern \& Harris
\cite{bir:ste}) and
thus some mechanism to supply fresh electrons in the hotspots is needed.
Strong shocks in the jet are considered to be the most probable
locations for particle acceleration. For instance, Falle \& Wilson
(\cite{fal:wil}) modelled the M 87 jet with a steady fluid jet to
which shocks are triggered by the decreasing pressure of the
interstellar medium. They interpret knot A in M 87 as location of a
strong shock where a reconfinement of the jet occurs. Similar 
mechanism could be working in the bright 3\arcsec\ knot in 3C 371,
although the appearance of both knots is more like a Mach disk
and not a conical shape like in the model of Falle \& Wilson 
(\cite{fal:wil}). 

The jet in 3C 371 apparently changes direction at the location
of the radio and optical knot . This property is shared only by 3C 346
that displays even more sudden bend at the location of a brigtening in
the optical jet (de Koff et al. \cite{kof:bau}).
If the jet in 3C 371 is relativistic at 
the projected distance of 3\arcsec\ from the nucleus then differences 
in the Doppler factor along the jet could produce apparent
brightenings. An enhancement of a factor of $\geq$ 10 is needed in
the R-band to raise the 3\arcsec\ knot from the background to the 
observed level. Wrobel \& Lind (\cite{wro:lin}) concluded that
the viewing angle $\theta$ is moderate for 3C 371. If we
assume $\theta$ = 30\arcdeg\ and a Lorenz factor $\gamma$ = 4 in the
jet, then a 5\arcdeg\ change in the jet direction towards the
observer is sufficient. For $\gamma$ = 2 and $\gamma$ = 1.5 the
corresponding values are 10\arcdeg\ and 20\arcdeg, respectively.  
Relativistic electrons moving with bulk relativistic speed in a
helical jet
is one possible scenario in this case. The structure seen in the
radio map of Akujor et al. (\cite{aku:lud}) lends some support to this
scenario, but it may not be
realistic since it is not known if the electrons really move with 
bulk relativistic speeds in jets at a distance of $\sim$ 4 kpc from the 
nucleus. 

The change of direction could also be attributed to a pressure
increase in the interstellar medium that diverts the jet from its
original direction. This kind of redirection of the jet has been 
suggested to be the reason for double hotspots in the edge-brightened
double radio sources (Williams \& Gull \cite{wil:gu1}, \cite{wil:gu2}).
The pressure increase could be supplied by the interaction of the
3C 371 host with neighboring galaxies. Figure 4 shows the 3C 371
field in the R-band after a smooth de Vaucouleurs model galaxy
has been subtracted from the image. A forced de Vaucouleurs profile
was used here to bring out the deviations from a smooth profile more
clearly. The model was fit to the
inner r $<$ 13\arcsec\ region which is the region reasonably well 
represented by de Vaucouleurs profile. Figure 4 clearly shows the
bridge of emission connecting the 3C 371 host to two spiral galaxies in the
SW corner of the field as noted earlier by Stickel, Fried \& K\"uhr 
(\cite{sti:fri}). Their spectroscopy also shows that the brighter
of these two galaxies lies at the same redshift as the 3C 371 host.
The bridge seems to continue to a distance
of $\sim$ 20 kpc from the nucleus
of the 3C 371 host. We do not see the bridge to extend closer to the
nucleus than this, but it is still tempting to hypothesize that
gas drawn from the nearby spiral galaxies is responsible for the turn
in the jet. Evidence for nonsymmetrical distribution of gas around 3C
371 nucleus was presented by Stickel, Fried \& K\"uhr (\cite{sti:fri})
who showed that the 
[\ion{O}{3}] $\lambda \lambda$ 4959,5007 emission is more extended
towards the companion galaxies in the SW than towards the opposite direction.
They interpreted this as a sign of tidal interaction between
the 3C 371 host and the SW galaxies.

Evidence for jet interaction with the environment has been
presented by previous authors both in smaller and larger scales than
in 3C 371. Baum et al. (\cite{bau:dea})  presented the results of 
a HST and MERLIN study of 3C 264 which also exhibits an optical jet.
In 3C 264 there is strong evidence for interaction between the jet and
a high density circumnuclear region at 300 - 400 pc radius. 
Also, evidence for strong jet-cloud interaction in the quasar 3C 254
at a projected distance of $\sim$ 15 kpc was presented by Crawford \& 
Vanderriest (\cite{cra:van}). The 4 kpc distance of interaction in 3C
371 is thus not unprecedented. It therefore seems, that the environmental
effects can have a significant impact on jet appearance,
particulary in the case of 3C 371 where the confinement of the jet 
probably plays a major role in the production of optical emission.   

\acknowledgements{The authors thank Joan Wrobel for making the 6 cm
radio image available. This work has been supported by the Academy of
Finland and by the Deutsche Forschungsgemeinschaft through
SFB 328 (J. Heidt) and FR 325/42-1 (K. J\"ager).}

\clearpage

\begin{deluxetable}{cccccc}
\footnotesize
\tablecaption{Log of observations}
\tablewidth{0pt}
\tablehead{
\colhead{Date} & \colhead{Telescope} & \colhead{Band} & 
\colhead{Detector} & \colhead{Exposure} & \colhead{Seeing}\nl
& & & & (s) & (arcsec) } 
\startdata
20 Oct 1995 & NOT             & R    & TEK 1024 $\times$ 1024   & 3600 & 0.83\nl
17 Jun 1996 & NOT             & B    & Loral 2048 $\times$ 2048 & 3600 & 0.80\nl
09 Jul 1996 & Calar Alto 3.5m & K$'$ & NICMOS3 256 $\times$ 256 & 1800 & 1.15\nl
13 Jul 1996 & NOT             & R    & TEK 1024 $\times$ 1024   & 1830 & 0.65\nl
\enddata
\end{deluxetable}

\begin{deluxetable}{cccc}
\footnotesize
\tablecaption{Photometry of the 3\arcsec\ knot.}
\tablewidth{0pt}
\tablehead{
\colhead{Band} & \colhead{dist}    & \colhead{PA}     &  \colhead{Flux}\nl
               & \colhead{\arcsec} & \colhead{\arcdeg} & \colhead{$\mu$Jy} } 
\startdata
B & 3.2 & 242 & 6.0 $\pm$ 0.6 \nl
R & 3.0 & 240 & 11 $\pm$ 1 \nl
K & 3.2 & 242 & 15$^{+7}_{-5}$ \nl 
\enddata
\end{deluxetable}

\newpage

\figcaption{A mosaic showing a 19\arcsec $\times$ 19\arcsec\ field
around the nucleus of 3C 371 after subtraction of the core and galaxy
light. The panels on the left show the ``raw'' images and the panels
on the right the same fields after 40 deconvolutions with the Lucy 
algorithm. Top row (a,b) shows the B-band image, middle row (c,d) the 
R-band image of 13 Jul 1996 and bottom row (e,f) the K$'$-band 
image. A small cross marks the position of the optical nucleus.
Light smoothing has been applied to the K$'$-band image. North is
up and east is to the left in all panels.}

\figcaption{The deconvolved R-band image of 13 Jul 1996 overlaid with
the 18 cm radio contours of Akujor et al. (1994), used here with
permission. The field size is 8\farcs3 $\times$ 7\farcs2, north is up
and east is to the left. The arrow indicates the direction of the
jet after the bright spot 3\arcsec\ from the nucleus. The tick marks
are at 0.5\arcsec\ intervals.}

\figcaption{The broadband spectrum of the knot 3\arcsec\ from the
nucleus. The errorbars are 2$\sigma$ errors. The fitted line has a
slope of -0.81. The radio data are from Browne et al. (1982) (408
MHz), Perley, Fomalont \& Johnston (1980) (1500 MHz), Perley \& 
Johnston (1979) (1480, 4900 MHz), Akujor et al. (1994) (1660 MHz) and
Pearson, Perley \& Readhead (1985) (4885 MHz). Optical and near
infrared data are from this paper.}

\figcaption{The R-band image of the field around 3C 371. The insert at
the lower left corner shows the image with low contrast. 
In the large image a galaxy model has been subtracted as described in
the text, the image has been smoothed lightly and it is shown here
with high contrast to 
emphasize low surface brightness areas. The optical image has been
overlaid by the countours of the 5 GHz radio image of Wrobel \& Lind
(1990). The field size is 176\arcsec $\times$ 159\arcsec\ and
north is up and east is to the left. Note the bridge of emission
connecting 3C 371 host to two spiral galaxies in the SW corner.}

\end{document}